\documentclass[prd,amsmath,amssymb,longbibliography,superscriptaddress,twocolumn,nofootinbib,10pt]{revtex4}

\pdfoutput=1

\usepackage{graphicx}
\usepackage{dcolumn}
\usepackage{bm}
\usepackage{amssymb}
\usepackage{latexsym}
\usepackage{booktabs}
\usepackage{amsmath}
\usepackage{multirow}

\usepackage[colorlinks=true, linkcolor=blue, citecolor=blue]{hyperref}

\newcommand{\LCDM}{\rm{\Lambda}CDM}
\newcommand{\Mpc}{\mathrm{~km~s^{-1}~Mpc^{-1}}}
\begin{document}

\title{Dynamical Dark Energy in the Crosshairs: A Joint Analysis with DESI, Pantheon plus, and TDCOSMO Constraints}

\author{Tong-hua Liu}
\email{liutongh@yangtzeu.edu.cn;}
\affiliation{School of Physics and Optoelectronic, Yangtze University, Jingzhou 434023, China}
\author{Xiaolei Li}
\email{lixiaolei@hebtu.edu.cn}
\affiliation{College of Physics, Hebei Normal University, Shijiazhuang 050024, China}
\author{Jieci Wang}
\email{jcwang@hunnu.edu.cn}
\affiliation{Department of Physics, and Collaborative Innovation Center for Quantum Effects and Applications, Hunan Normal University, Changsha 410081, China;}

\begin{abstract}
In this work, we perform a comprehensive joint analysis of three representative dark energy models - $\LCDM$, the Chevallier-Polarski-Linder (CPL) parametrization, and the Generalised Emergent Dark Energy (GEDE) model - using the latest observational datasets: baryon acoustic oscillation (BAO) measurements from DESI Data Releases 1 and 2 (DR1/DR2), the Pantheon Plus sample of Type Ia supernovae (SNe Ia), and time-delay cosmography from TDCOSMO lensing. The CPL model yields a statistically significant improvement over $\LCDM$, with $\Delta \chi^2 = -3.6$ for DESI DR2+Pantheon Plus+TDCOSMO, favoring a quintessence-like behavior ($w_0=-0.87^{+0.05 }_{-0.05} $, $w_a=-0.41^{+ 0.28}_{-0.28}$ at 1$\sigma$ confidence level).  The GEDE model also exhibits slightly favoring  than $\LCDM$ with observations, yielding $\Delta \chi^2 =-2.83$ (DR1) and $\Delta \chi^2 = -3.82$ (DR2). The captured potential dark energy evolution transition parameter $\Delta$ in GEDE model is constrained to $-0.52^{+0.22}_{-0.20}$ (DR1) and $-0.50^{+0.18}_{-0.17}$ (DR2), showing a $2.4\sigma$ and $2.9\sigma$ deviation from zero respectively. This statistically significant ($>2\sigma$) non-zero value of $\Delta$ provides evidence against a pure cosmological constant scenario. The negative sign indicates quintessence-like behavior ($w > -1$) in the late universe. Notably, the GEDE constraints show a $>3\sigma$ tension with the $\Delta=1$ (Phenomenological Emergent Dark Energy model) predictions. This significant discrepancy implies that while both models belong to the emergent dark energy family, their fundamentally different transition mechanisms lead to distinct cosmological implications. Our results collectively suggest that  GEDE provides a phenomenologically viable alternative to $\Lambda$CDM.

\end{abstract}

\maketitle

%\bigskip
\section{Introduction}

The discovery of cosmic acceleration through Type Ia supernovae (SNe Ia) \cite{1998AJ....116.1009R,1999ApJ...517..565P} marked a paradigm shift in modern cosmology, leading to the establishment of the standard  $\LCDM$ model. This framework, further reinforced by precision measurements of cosmic microwave background (CMB) anisotropies \cite{2020A&A...641A...6P} and large-scale structure \cite{2021PhRvD.103h3533A}, describes a universe dominated by cold dark matter (CDM) and a cosmological constant ($\Lambda$). However, as observational precision has improved, significant tensions have emerged between early- and late-Universe probes that challenge this standard picture. The most prominent discrepancy is the Hubble tension, over $4.4\sigma$ statistic difference between the Planck CMB inferred Hubble constant ($H_0$ = 67.4 $\pm$ 0.5 $\Mpc$) and local distance ladder measurements ($H_0$ = 73.04 $\pm$ 1.04$\Mpc$) \cite{2022ApJ...934L...7R}. Equally puzzling is the $S_8$ tension, where the Planck value of $\sigma_8(\Omega_m/0.3)^{0.5} = 0.832\pm 0.013$ diverges from weak lensing surveys like Kilo-Degree Survey  ($S_8=0.776 \pm 0.017 $) \cite{2021A&A...645A.104A}. These persistent discrepancies have motivated extensive investigations into extensions of $\Lambda$, including dynamical dark energy models and modified gravity scenarios.
More discussions on Hubble tension please see the references \citep{2019PhRvL.122v1301P,2019PhRvL.122f1105F,2021ApJ...912..150D,2020ApJ...895L..29L,
2023arXiv230306974D,2022ApJ...939...37L,2023ApJS..264...46L,2020PhRvD.102b3518V,2023Univ....9..393V,2021PhRvD.103h1305B,2022JCAP...04..004L} and  references therein for a more comprehensive discussion.

The Dark Energy Spectroscopic Instrument (DESI), currently completing its 5-year survey, has emerged as a powerful tool to address these challenges. Its Data Release 1 (DR1) \cite{2025JCAP...02..021A} and Data Release 2 (DR2) \cite{2025arXiv250314738D} have delivered ground-breaking results from spectroscopic observations of over 40 million galaxies and quasars, providing unprecedented baryon acoustic oscillation (BAO) measurements across $0.1 < z < 4.2$. While DR1 achieved sub-percent precision on Hubble parameter and angular diameter distance, DR2 has further improved these constraints by incorporating additional sky coverage and improved redshift measurements.
Recent analyses of the DESI DR2 have significantly advanced our understanding of dynamical dark energy. By combining BAO measurements from 14 million galaxies and quasars with CMB and SNe Ia data, these studies reveal a  $2.8-4.2\sigma$ preference for evolving dark energy ($w_0 > -1, w_a < 0$) over the standard  $\LCDM$.  This preference is further supported by correlated evidence from Lyman-$\alpha$ forest BAO measurements and neutrino mass constraints, where $\sum m_\nu<0.064$ eV under $\Lambda$CDM  and $\sum m_\nu<0.16$ eV in assuming  the $w_0w_a$CDM model \cite{2025arXiv250314738D}. These findings collectively challenge the viability of  $\LCDM$ and align with independent evidence for redshift-evolving dark energy. Notably, both shape-function reconstruction and non-parametric methods incorporating a correlation prior derived from Horndeski theory consistently corroborate this dynamical behavior \cite{2025arXiv250406118G}. Beyond dark energy, the galaxy clustering measurements from DESI DR2 reveal that its $S_8$ value exhibits better agreement with Planck CMB data compared to previous optical surveys (e.g., Kilo-Degree Survey \cite{2021A&A...645A.104A}, Dark Energy Survey \cite{2018PhRvD..98d3526A}) \cite{2025arXiv250314738D}. This result may partially alleviate the tension between CMB and weak gravitational lensing observations, though further validation of its statistical significance remains necessary and it highlights the crucial role of DESI data in testing cosmological models beyond  $\LCDM$.  For a non-exhaustive set of references on this topic, see \cite{2025arXiv250400985Y,2025arXiv250321652S,2025arXiv250319898P,2025arXiv250319352I,2024ApJ...976....1L,2025arXiv250314743L,
2025arXiv250204212H,2025arXiv250407679W,2025arXiv250416868A,2025arXiv250310806C,2024PhRvD.110l3533P,2024arXiv241013627P,2025arXiv250103480P,2022PhRvD.106e5014Y,2024JHEP...05..327Y,2025arXiv250318417L,2024PhRvD.110l3519J,2025JCAP...01..153J,2011PhLB..698..175C,2025arXiv250404646M,2025arXiv250404417C,2025arXiv250324343C,2025arXiv250219274W,2025arXiv250321600P,2025arXiv250322529N,2025arXiv250323225S}.

To fully exploit the potential of DESI data and conduct in-depth research on the dynamics of dark energy, we systematically investigated three representative cosmological models: the standard $\LCDM$ model as the benchmark framework, the dynamically evolving dark energy scenario described by the Chevallier-Polarski-Linder (CPL) parametrization \cite{2001IJMPD..10..213C,2003PhRvL..90i1301L}, and  the novel Generalised Emergent Dark Energy (GEDE) model where quantifies the dark energy density flexibility and generality on the behaviour of dark energy evolution \cite{2020ApJ...902...58L}. By employing a multi-probe joint analysis approach that integrates BAO measurements from DESI DR1/DR2,  Pantheon Plus sample \cite{2022ApJ...938..113S} which consists of 1701 light curves of 1550 distinct SNe Ia ranging in redshift from $0.001<z<2.261$, and Time-Delay COSMOgraphy (TDCOSMO) gravitational lensing time-delay observations \cite{2020A&A...643A.165B}, this synergistic approach effectively breaks key parameter degeneracies, particularly between $H_0$, the sound horizon $r_d$, and the supernova absolute magnitude $M_B$, while providing multiple independent constraints on cosmic expansion. Notably, TDCOSMO's time-delay cosmography yields a high precision $H_0$ measurement that is independent of both Cepheid distance ladder and CMB methods, offering unique leverage on the Hubble tension and remaining unaffected by systematics plaguing other probes.

This paper is organized as follows:  Section 2 outlines the observational data and analysis methodology. We present and discuss the results of our statistical analysis. Finally, the main conclusions are summarized in Section 4.
%2019ApJ...883L...3L

\section{ Data and Methodology}\label{sec:data}
We outline the dark energy models including $\LCDM$ model, GEDE model, CPL model, and  parameter sampling, as well as the priors on parameters in this Section, and presents a discussion on the cosmological datasets utilized in this study.

The mysterious driver of cosmic acceleration, dark energy, is phenomenologically modeled as a fluid with equation of state $p= w\rho$ ($p$ denotes pressure and $\rho$ represents energy density), where the simplest $\LCDM$  scenario $w = -1$ and $\Omega_{DE}=(1-\Omega_{m})=$constant in a flat universe, where $\Omega_m$ is the matter density at present time.

The critical question of whether $w$ evolves with redshift, which would carry profound implications for fundamental physics, motivates our investigation of observationally tractable parameterizations. To investigate the dynamical evolution of dark energy, several phenomenological parameterizations have been proposed. The simplest phenomenology could be just a first-order Taylor expansion with redshift or scale factor,  such as  widely used CPL parametrization \cite{2001IJMPD..10..213C,2003PhRvL..90i1301L}, the equation of state of dark energy is given by $w(z)=w_0+w_az/(1+z)$, and further derive $\Omega_{DE}=(1-\Omega_{m})\times(1+z)^{3(1+w_0+w_a)}\exp(\frac{-3w_az}{1+z})$. This parametrization has several advantages such as well-bounded behavior at high redshifts, simple physical interpretations, and manageable two-dimensional phase space.

Motivated by the current status of the cosmological observations and significant tensions in the estimated values of some key parameters assuming the standard $\LCDM$ model, \citet{2020ApJ...902...58L}  proposed a simple but radical emergent dark energy model in light of a series of cosmological probes and considering the evolution of the model at the level of linear perturbations. In GEDE model, the dark energy density has the following form $\Omega_{DE}=(1-\Omega_{m})\frac{[1-\Delta\tanh(\log_{10}({1+z}/{1+z_t}))]}{[1+\Delta\tanh(\log_{10}(1+z_t))]}$,  and the equation of state of dark energy has $w(z)=-\Delta/(3\ln 10)\times[1+\tanh(\Delta\log_{10}(\frac{1+z}{1+z_t}))]-1$, where $\Delta$ is the only free parameter which can take any
real values, and $z_t$ denotes the epoch where the matter energy density and the dark energy density are equal, that
means $\rho_m(z_t)=\rho_{DE}(z_t)$. Thus, $z_t$ is a derived parameter, not a free parameter, please see work \cite{2020ApJ...902...58L} for more detail on this model. When $\Delta=0$, this model reverts to the $\LCDM$ model. When $\Delta=1$, the model transforms into the Phenomenological Emergent Dark Energy (PEDE) model,  where dark energy is absent at early Universe. As cosmic time progresses, dark energy emerges gradually.  This model was proposed by \citet{2019ApJ...883L...3L}, and  has garnered significant attention for its novel description of dark energy dynamics. Importantly, PEDE model can significantly reduce the Hubble tension in estimation of the cosmological parameters using low- and high-redshift data, some level of tension remains. However, current DESI DR2 findings suggest that the PEDE model, whose equation of state  asymptotically approaches $w=-1$, may be insufficient to describe the observed dark energy dynamics. In this study, we instead employ the GEDE model, which not only encompasses all features of PEDE but also introduces critical flexibility through its $\Delta$ parameter: a positive $\Delta$ yields a phantom-like equation of state  ($w<-1$), while a negative $\Delta$ produces quintessence-like behavior ($w>-1$). This parametric control allows GEDE to more comprehensively capture potential dark energy evolution scenarios suggested by modern observational constraints. Recently work on GEDE and PEDE models please see refs \cite{2024MNRAS.533.1865L,2021PDU....3100762Y,2022A&A...668A..51L,2024MNRAS.52711962E,2024EPJC...84..444L}.

The datasets used in the analyses are described below:

$\bullet$ $\textbf{DESI DR1 and DR2:}$
We employ the latest BAO measurements from the DESI collaboration, which combine observations of galaxies, quasars, and Lyman-$\alpha$ tracers to provide robust constraints on cosmological distances - specifically the  transverse comoving distance $D_M/r_d$, the Hubble horizon $D_H/r_d$, and the angle-averaged distance $D_V/r_d$, all normalized to the comoving sound horizon at the drag epoch $r_d$, and reported in Table I of the first data release (DR1) \cite{2025JCAP...02..021A} and Table IV of the second release (DR2) \cite{2025arXiv250314738D}. Our analysis fully incorporates the correlation structure of these measurements through the cross-correlation coefficients $r_{M,H}$ between $D_M/r_d$ and $D_H/r_d$, ensuring proper treatment of the covariance between different distance scales. In what follows, we make use of the DESI DR1 and DESI DR2 datasets separately, without combining them.

$\bullet$ $\textbf{Pantheon Plus}$:
The Pantheon Plus sample \cite{2022ApJ...938..113S} includes 1701 light curves from 1550 distinct SN Ia events, spanning a redshift range of 0.01 to 2.26.
In order to reduce the impact of calibration systematics on cosmology, we therefore does not include the calibration sample at lower redshifts, and take into account covariance matrix $\footnote{\url{https://github.com/PantheonPlusSH0ES/DataRelease}}$.

$\bullet$ $\textbf{TDCOSMO Lenes}$: We analyze a combined sample of 7 time-delay lenses from the TDCOSMO collaboration $\footnote{\url{http://tdcosmo.org}}$ and 33 additional Sloan Lens ACS (SLACS) systems, adopting the hierarchical Bayesian approach developed by the TDCOSMO team to address the dominant mass-sheet transformation (MST) systematic uncertainty in time-delay cosmography, where stellar kinematics constraints from the full lenses sample help break the MST degeneracy while incorporating both imaging and spectroscopic data  \cite{2020A&A...643A.165B}. Here, we use the 7 time-delay  TDCOSMO lenses  incorporated integral field unit (IFU) spectroscopy from VIMOS 2D data of a subset of the SLACS lenses.

Recently, \citet{2024MNRAS.529L..95H} publicly released an external likelihood package$\footnote{\url{https://github.com/nataliehogg/tdcosmo_ext}}$ for the cosmological modeling and sampling software \cite{2021JCAP...05..057T}. This package enables the use of the hierarchical TDCOSMO likelihood to constrain cosmological model parameters, allowing for seamless integration with other cosmological likelihoods and Boltzmann code configurations. In our analysis, we incorporate this code and combine it with the likelihoods from DESI BAO and SNe Ia to study our target cosmological models.
To ensure robust convergence of the Markov Chain Monte Carlo (MCMC) chains, we apply the Gelman-Rubin diagnostic criterion \cite{1992StaSc...7..457G}. For all models and datasets, we impose a strict convergence threshold of $R-1<0.01$  before accepting the results.
\begin{figure}
{\includegraphics[width=1\linewidth]{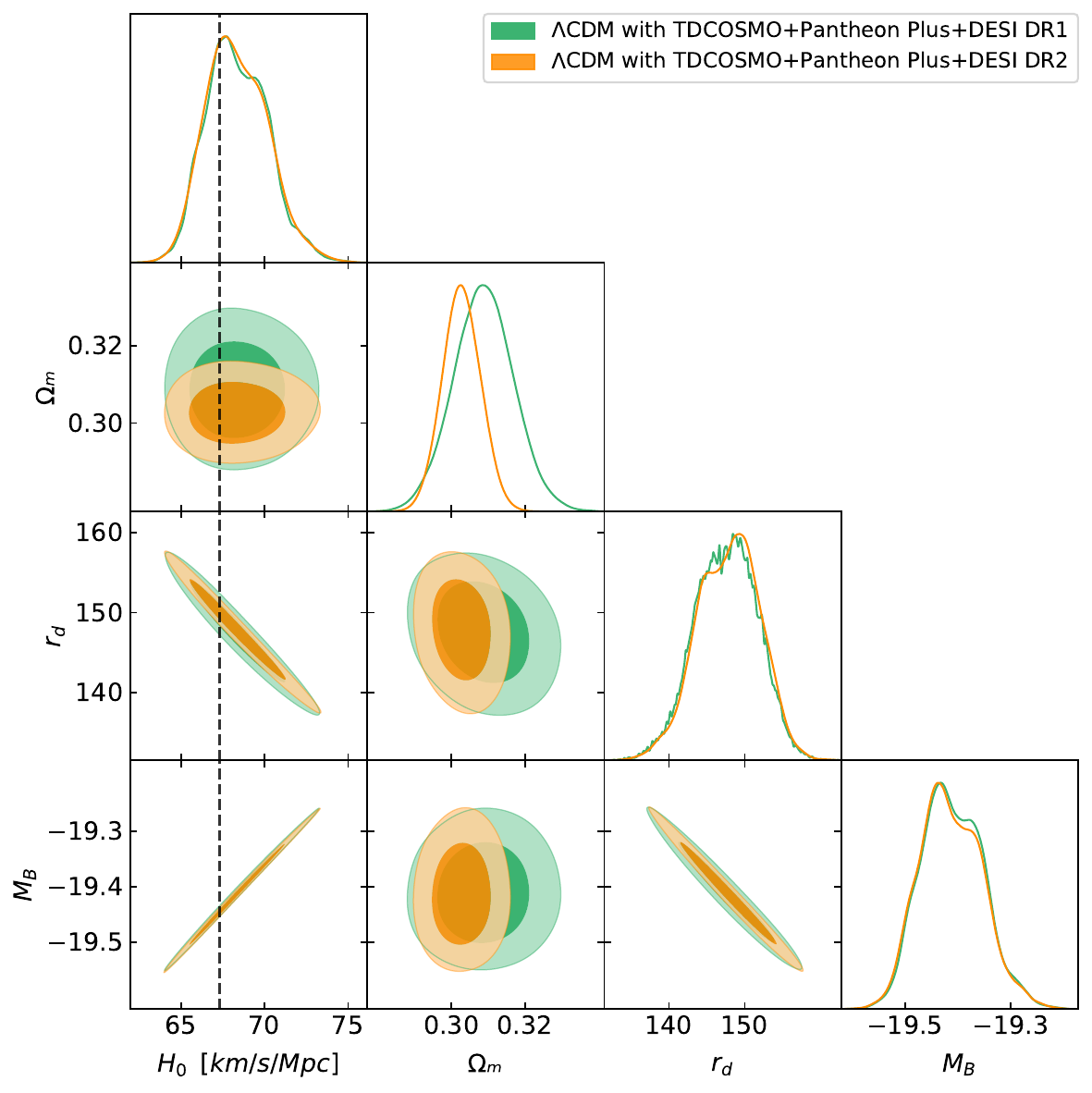}}
\caption{The one and two-dimensional marginalised posterior distributions of $H_0$, $\Omega_m$, $r_d$, $M_B$ in a $\LCDM$ cosmology from the DESI DR1/DR2+Pantheon Plus+TDCOSMO lenes datasets (the green regions derived from DESI DR1 dataset, the yellow regions derived from DESI DR2 dataset). The gray dashed line is the best-fitting value of $H_0=67.4$ $\Mpc$ from Planck CMB results.}
\label{LCDM}
\end{figure}

We adopt flat priors on the sampled cosmological parameters $\{H_0, \Omega_m, r_d, M_B, w_0, w_a\}$ when considering the CPL model. For the $\LCDM$  model, the dynamical dark energy parameters $w_0$ and $w_a$  are omitted. For the GEDE model, the sampled cosmological parameters is $\{H_0, \Omega_m, r_d, M_B, \Delta\}$.  The resulting Markov Chain Monte Carlo (MCMC) chains are post-processed using the \textbf{GetDist} package, which facilitates the computation of numerical constraints, including 1-D posteriors and 2-D marginalized probability contours.
\begin{table*}
\renewcommand\arraystretch{2.5}
\caption{\label{tab1} Summary of the constraints on the the parameters in the framework of $\LCDM$, GEDE and CPL models by using TDCOSMO lensing (denotes TDC), Pantheon Plus (denotes SN), and DESI DR1/DR2 observations.}
\centering
\scalebox{0.9}{
\begin{tabular}{ l| c c c c c c c c}

\hline
\hline
Cosmological Model with Dataset  & $H_0 $ &$\Omega_m$ & $r_d$  & $M_B$ & $w_0$ & $w_a$  & $\Delta$ & $\chi^2$ %&$\nu$
\\
\hline
$\LCDM$ TDC+SN+DESI DR1 \,\, & \,\,$68.25^{+1.97}_{-1.73}$ & $0.308^{+ 0.008}_{-0.008}$ & $ 147.64^{+4.03}_{-4.23}$& $-19.414^{+0.062}_{-0.056}$  & $-$  & $-$& $-$&1449.38\\
\hline

$\LCDM$ TDC+SN+DESI DR2   & \,\,$ 68.21^{+ 2.02}_{-1.72}$ & $0.302^{+ 0.005}_{-0.005}$ & $ 148.05^{+ 3.90}_{-4.30}$& $-19.417^{+0.064}_{-0.056}$  & $-$ & $-$& $-$&1449.88\\
\hline
GEDE TDC+SN+DESI DR1   & \,\,$ 67.58 ^{+ 2.03}_{-1.69} $& $0.297^{+0.010}_{-0.010}$ & $ 148.21^{+3.95}_{-4.30}$& $-19.426^{+0.064}_{ -0.054}$  & $-$ & $-$ & $-0.52^{+0.22}_{-0.20}$&1446.06\\
\hline
GEDE TDC+SN+DESI DR2  & \,\,$  67.62^{+ 2.00}_{-1.68} $& $ 0.298^{+0.006}_{ -0.005}$ & $147.69^{+3.74}_{-4.24}$& $-19.425^{+ 0.063}_{ -0.054}$  & $-$ & $-$& $-0.50^{+0.18}_{-0.17}$&1446.55\\
\hline
CPL TDC+SN+DESI DR1 & $ \,\,67.24^{+ 2.03}_{-1.76} $& $ 0.311^{+0.012 }_{-0.016}$& $  149.04^{+ 4.11}_{-4.41}$& $ -19.432^{+0.064}_{-0.056}$  & $ -0.87^{+0.06 }_{-0.05}$ & $ -0.53^{+ 0.45 }_{-0.43}$  & $-$ &1445.23\\
\hline
CPL TDC+SN+DESI DR2    & \,\,$ 67.33^{+2.05}_{-1.79} $& $ 0.310^{+ 0.009 }_{-0.010}$ & $  148.00^{+ 4.07}_{ -4.32}$& $ -19.428^{+0.064}_{-0.057}$  & $ -0.87^{+0.05 }_{-0.05}$ & $ -0.41^{+ 0.28 }_{-0.28}$ & $-$&1445.77\\
\hline
\hline
\end{tabular}
}
\end{table*}

\section{Results and Discussion}
\begin{figure}
{\includegraphics[width=1\linewidth]{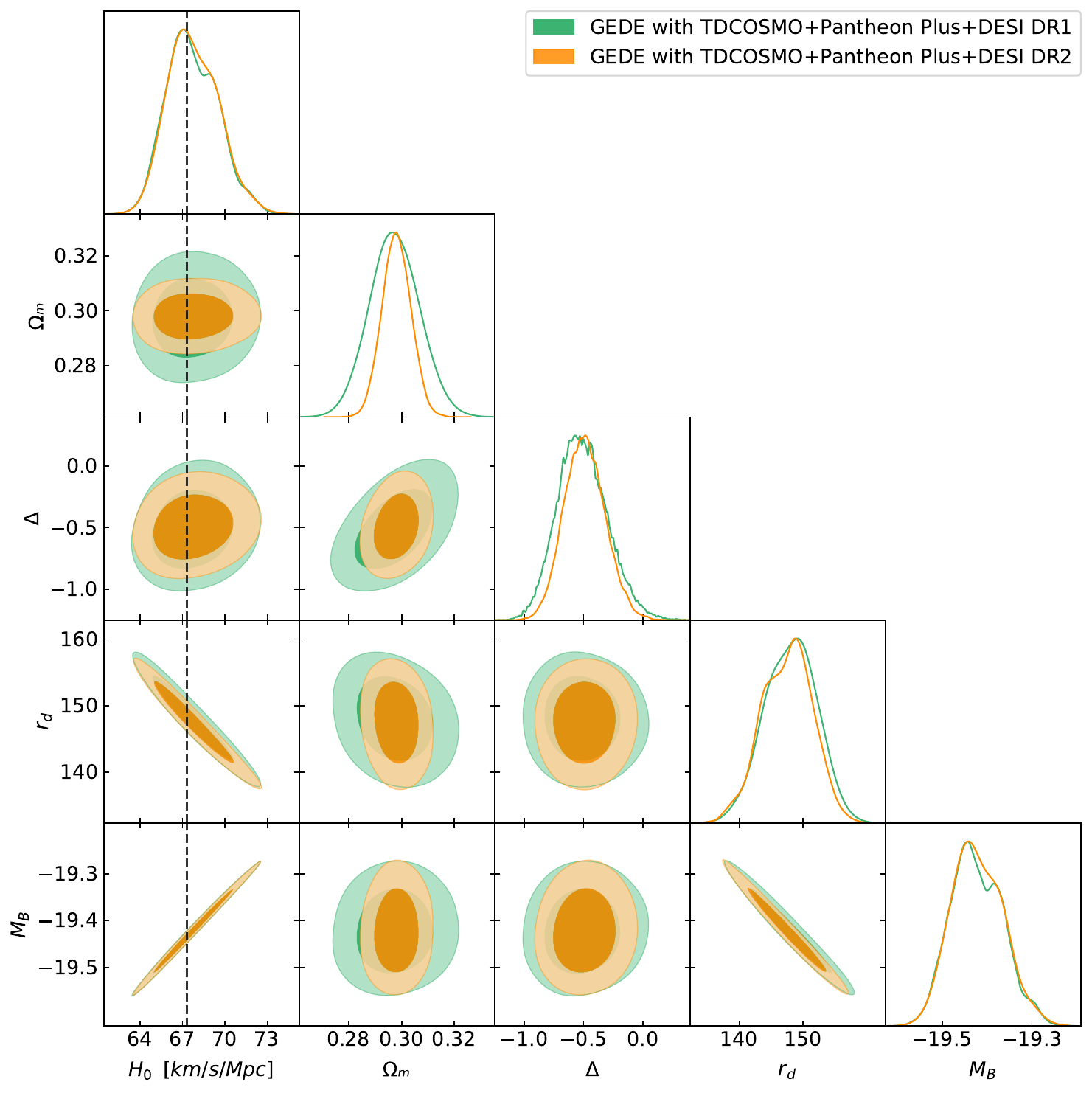}}
\caption{The one and two-dimensional marginalised posterior distributions of $H_0$, $\Omega_m$, $\Delta$, $r_d$, $M_B$ in a GEDE cosmology.}
\label{GEDE}
\end{figure}

We restrict our analysis to a spatially flat universe. Our investigation begins with the $\LCDM$ model, incorporating cosmological constraints from DESI DR1/DR2, Pantheon Plus SNe Ia, and TDCOSMO lensing datasets. Figure \ref{LCDM} presents the 1D marginalized posterior distributions and 2D joint confidence regions at the
$1\sigma$ and $2\sigma$  confidence levels for key cosmological parameters, where the dark and light green shaded regions correspond to constraints derived from the DESI DR1+Pantheon Plus+TDCOSMO lensing combination. For the DESI DR1 dataset, we obtain the following best-fit values with $1\sigma$  uncertainties: $H_0 =68.25^{+1.97}_{-1.73} \Mpc$,  $\Omega_m=0.308^{+ 0.008}_{-0.008}$, $r_d=147.64^{+4.03}_{-4.23}$ Mpc and $M_B=-19.414^{+0.062}_{-0.056}$ mag. These results show excellent agreement with both Planck 2018 CMB observations \cite{2020A&A...641A...6P} and TDCOSMO lensing constraints \cite{2020A&A...643A.165B}. The DESI DR2 analysis yields consistent but slightly improved constraints: $H_0 = 68.21^{+ 2.02}_{-1.72} \Mpc$,  $\Omega_m=0.302^{+ 0.005}_{-0.005}$, $r_d=148.05^{+ 3.90}_{-4.30}$ Mpc and $M_B=-19.417^{+0.064}_{-0.056}$ mag, and also showed in Figure \ref{LCDM} with the dark and light yellow shaded regions.  Notably, while DR2 provides a $\sim 40\%$ tighter constraint on $\Omega_m$ compared to DR1 (uncertainty reduced from 0.008 to 0.005) other parameters remain statistically consistent between the two data releases. The absolute magnitude $M_B$ values cluster around
-19.4 mag in remarkable agreement with the inverse distance ladder measurement of  $M_B=-19.396\pm0.016$ reported by \citet{2023PhRvD.107f3513D} using Pantheon SNe Ia+BAO datasets. This consistency across independent methods and datasets reinforces the robustness of our results. The complete numerical constraints are summarized in Table \ref{tab1}.

\begin{figure*}
{\includegraphics[width=0.9\linewidth]{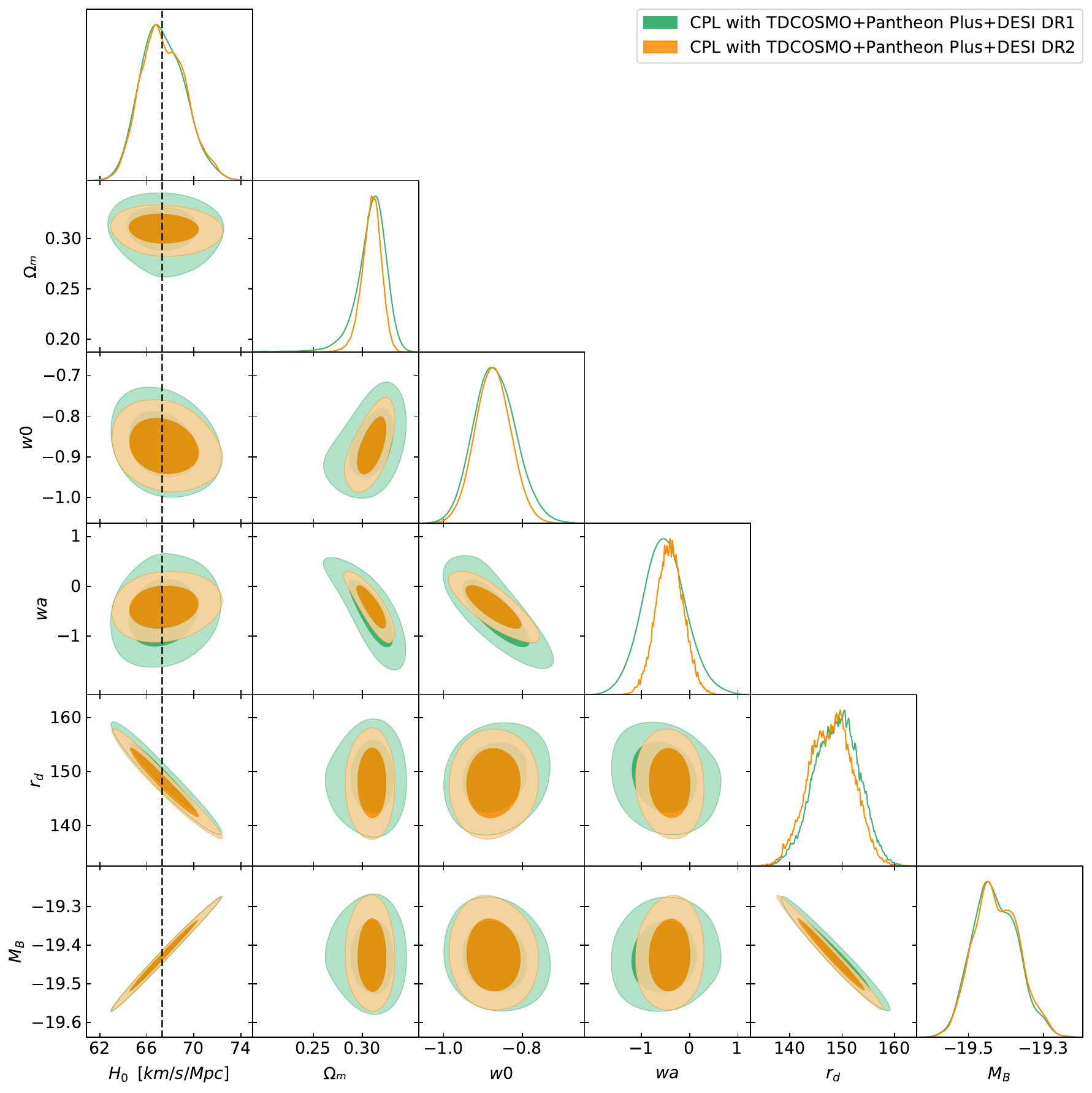}}
\caption{The one and two-dimensional marginalised posterior distributions of $H_0$, $\Omega_m$, $w_0$, $w_a$, $r_d$, $M_B$ in a CPL cosmology from the DESI DR1/DR2+Pantheon Plus+TDCOSMO lenses datasets (the green regions derived from DESI DR1 dataset, the yellow regions derived from DESI DR2 dataset). The gray dashed line is the best-fitting value of $H_0=67.4$ $\Mpc$ from Planck CMB results.}
\label{CPL}
\end{figure*}
Figure \ref{GEDE} presents the $1\sigma$ and $2\sigma$ confidence contours for cosmological parameters in the GEDE framework, with corresponding numerical constraints detailed in Table \ref{tab1}. The analysis reveals several key findings: The GEDE model yields $H_0 = 67.58^{+2.03}_{-1.69}$ km s$^{-1}$ Mpc$^{-1}$ (DESI DR1) and $67.62^{+2.00}_{-1.68}$ km s$^{-1}$ Mpc$^{-1}$ (DESI DR2), showing closer alignment with Planck CMB measurements than the $\Lambda$CDM predictions. The matter density parameter $\Omega_{{m}} = 0.297^{+0.010}_{-0.010}$ (DR1) and $0.298^{+0.006}_{-0.005}$ (DR2) remains remarkably consistent between data releases.
The transition parameter $\Delta$ is constrained to $-0.52^{+0.22}_{-0.20}$ (DR1) and $-0.50^{+0.18}_{-0.17}$ (DR2), showing a $2.4\sigma$ and $2.9\sigma$ deviation from zero respectively. This statistically significant ($>2\sigma$) non-zero value of $\Delta$ provides evidence against a pure cosmological constant scenario. The negative sign indicates quintessence-like behavior ($w > -1$) in the late universe, with the transition redshift $z_t \approx 0.36$ consistent across both datasets. Notably, the GEDE constraints show a $>3\sigma$ tension with the PEDE model predictions. This significant discrepancy implies that while both models belong to the emergent dark energy family, their fundamentally different transition mechanisms lead to distinct cosmological implications. The consistency between DESI DR1 and DR2 results, with variations smaller than $0.5\sigma$ for all parameters, demonstrates GEDE's stability against observational systematics. These results demonstrate the stability of the GEDE framework across different datasets while providing new insights into the dark energy, with the model naturally interpolating between early and late cosmological probes through its $\Delta$-parameterized transition mechanism. The observed consistency in $\Omega_{m}$ and $\Delta$ constraints suggests GEDE's robustness against observational systematics in current BAO measurements.

\begin{figure}
{\includegraphics[width=1\linewidth]{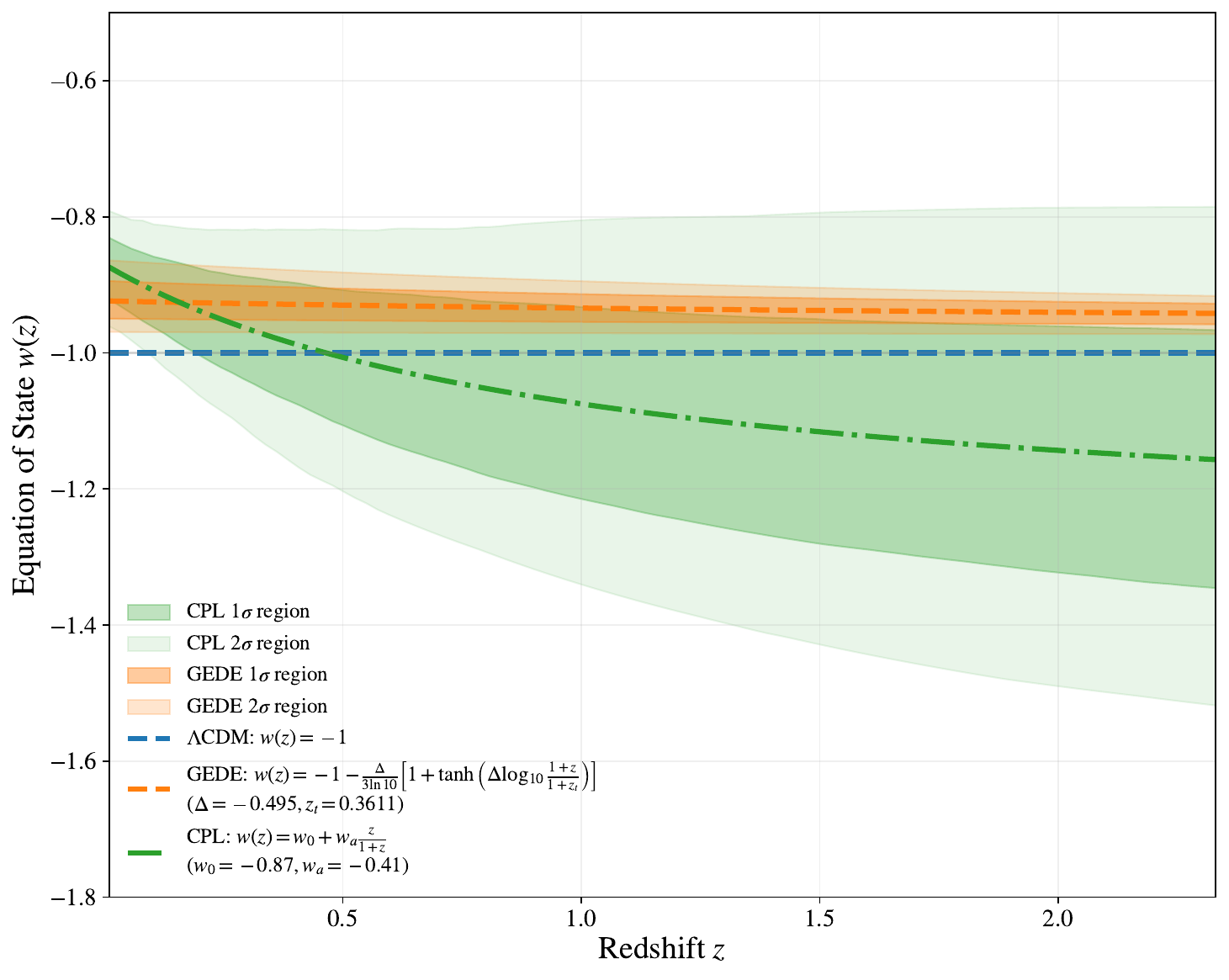}}
\caption{The equation of state of dark energy with $\LCDM$, GEDE, and CPL models using DESI DR2, Pantheon Plus SNe Ia, and TDCOSMO lensing datasets.  The dashed green line shows the best-fitting $w(z)$ based on $w_0$ and $w_a$ inference and the green regions around it represent the 68\% and 95\% confdence intervals. The dashed blue line represents  the  $\LCDM$ model, and  the dashed orange line represents GEDE model.}
\label{evlo}
\end{figure}
Within the CPL framework, our results corroborate the dynamical dark energy behavior reported in the recent DESI collaboration study \cite{2025arXiv250314738D} and other work \cite{2024JCAP...10..035G}. The combined DESI+Pantheon Plus+TDCOSMO lensing analysis yields compelling evidence for quintessence-like dark energy, with constraints on the present-day equation of state parameter of $w_0=-0.87^{+0.055 }_{-0.051} $ (DESI DR1) and  $w_0=-0.87^{+0.045 }_{-0.045} $ (DESI DR2) at 68\% confidence level - a statistically significant deviation (2.4$\sigma$ for DR1, 2.9$\sigma$ for DR2) from the cosmological constant value $w_0=-1$. The time evolution parameter shows mild but non-negligible dynamics, with $w_a=-0.53^{+ 0.45 }_{-0.43}$ (DR1) and  $w_a=-0.41^{+ 0.28 }_{-0.28}$ (DR2) representing 1.2$\sigma$ and 1.8$\sigma$ departures from zero respectively. These results suggest an evolving dark energy component that: i) currently resides in the quintessence regime ($w_0>-1$); ii) exhibits a clear evolutionary trend toward phantom values ($w(z)>-1$) at low redshift transitioning to $w(z)<-1$ at higher redshift; The improved constraints from DESI DR2 data strengthen the case for dynamical dark energy compared to previous analyses, reducing the uncertainty on $w_a $ by nearly 40\% while maintaining consistent central values.  The 1D marginalized posterior distributions and 2D joint confidence regions displays in Figure \ref{CPL} and numerical results of the constraints are shown in Table  \ref{tab1}.
The results of the dark energy evolution with redshift of these three models are shown in  Figure \ref{evlo}. We can obviously find that $w(z)$ crosses the value of -1 for CPL model by using the best-fitting $w(z)$ based on $w_0$ and $w_a$ inference from  DESI DR2, Pantheon Plus SNe Ia, and TDCOSMO lensing datasets.

It is particularly noteworthy that our analysis utilizing the combined DESI DR1/DR2, Pantheon Plus, and TDCOSMO lensing datasets achieves constraints on dark energy and its evolution parameters that are comparable in precision to those obtained in previous work \cite{2024JCAP...10..035G} using Planck CMB 2018 observations combined with SNe Ia and DESI data, despite yielding relatively less stringent constraints on the  $H_0$ and  $r_d$ parameters. The TDCOSMO lensing data play a pivotal role in this analysis by providing absolute distances that effectively anchor the relative distances derived from BAO and SNe Ia measurements, thereby breaking the strong degeneracy between the $H_0$, $r_d$ and $M_B$. Remarkably, our results demonstrate that low-redshift observations can now achieve precision comparable to Planck CMB measurements for dynamical dark energy studies, while offering complementary advantages in probing the late-time universe. This synergy between different observational data significantly enhances our ability to constrain dark energy properties.

The $\chi^2$ analysis reveals nuanced differences between cosmological models. For the DESI DR1+Pantheon Plus+TDCOSMO dataset, the GEDE model yields $\chi^2 = 1446.06$, showing a modest improvement ($\Delta\chi^2 \approx -3.3$) compared to the $\Lambda$CDM value of $1449.38$. This trend persists in the DESI DR2 combination, where GEDE achieves $\chi^2 = 1446.55$ versus $\Lambda$CDM's $1449.88$ ($\Delta\chi^2 \approx -3.3$). 
The CPL parametrization shows slightly better fits, with $\chi^2 = 1445.23$ (DR1) and $1445.77$ (DR2), corresponding to $\Delta\chi^2 \approx -4.2$ and $-4.1$ respectively relative to $\Lambda$CDM. While GEDE demonstrates comparable performance to CPL, both models show only marginal improvements over $\Lambda$CDM, with $\Delta\chi^2$ values within $-3$ to $-4$ across datasets. The consistent $\chi^2$ reductions suggest that while dynamical dark energy models may provide slightly better fits, the statistical evidence remains inconclusive given the small $\Delta\chi^2$ magnitudes.

\section{Conclusion}

The joint analysis of DESI DR1/DR2, Pantheon Plus, and TDCOSMO datasets provides important constraints on dynamical dark energy models. For the GEDE framework, we find $\chi^2$ values of $1446.06$ (DESI DR1) and $1446.55$ (DESI DR2), representing modest improvements of $\Delta\chi^2 \approx -3.3$ relative to the $\Lambda$CDM baseline in both cases. The key parameter $\Delta$ governing dark energy evolution is constrained to $-0.52^{+0.22}_{-0.20}$ (DR1) and $-0.50^{+0.18}_{-0.17}$ (DR2) at 68\% confidence level, with negative values consistently indicating quintessence-like behavior ($w > -1$) in the late universe.
Focusing on the CPL parametrization, we observe slightly better fits with $\Delta\chi^2 \approx -4.2$ and $-4.1$ relative to the $\Lambda$CDM baseline by using DESI DR1/DR2, respectively. The CPL parameters show $w_0 = -0.87^{+0.06}_{-0.05}$ and $w_a = -0.53^{+0.45}_{-0.43}$ for DR1, with DESI DR2 providing tighter constraints of $w_0 = -0.87^{+0.05}_{-0.05}$ and $w_a = -0.41^{+0.28}_{-0.28}$. These represent approximately 1-2$\sigma$ deviations from the $\Lambda$CDM expectation of $(w_0,w_a)=(-1,0)$, suggesting mild but non-negligible dark energy dynamics.

The multi-probe methodology proves particularly valuable in breaking degeneracies between cosmological parameters. By combining DESI BAO measurements ($r_d = 147.64^{+4.03}_{-4.23}$ Mpc for DR1), Pantheon Plus supernovae ($M_B = -19.414^{+0.062}_{-0.056}$ mag), and TDCOSMO time-delay lensing constraints, we achieve robust measurements of $H_0$ ($67.58^{+2.03}_{-1.69}$ km s$^{-1}$ Mpc$^{-1}$ for GEDE+DR1) that are consistent across different probes. The inferred supernova absolute magnitude remains stable at $M_B \approx -19.42$ mag in all analyses.

Our results collectively suggest that while GEDE provides a phenomenologically viable alternative to $\Lambda$CDM, the precise nature of dark energy evolution remains an open question.
Looking ahead, future studies should prioritize several directions: i) investigating systematic uncertainties in DESI BAO and SNe Ia calibrations; ii) exploring higher-order dark energy parametrizations or model-independent reconstructions; and iii) integrating upcoming datasets (e.g., CMB, Euclid, and Rubin Observatory) to test the redshift-dependence of dark energy across cosmic epochs. 
The persistent tensions between early- and late-Universe probes demand a concerted effort to refine both observational techniques and theoretical frameworks. Our findings solidify the role of multi-messenger cosmology in advancing our understanding of dark energy.

\section*{Acknowledgments}
This work was supported by the National Natural Science Foundation of China under Grants No. 12203009, No. 12475051, and No. 12035005; The National Key Research and Development Program of China (No. 2024YFC2207400); The Chutian Scholars Program in Hubei Province (X2023007); The Innovative Research Group of Hunan Province under Grant No. 2024JJ1006; The Science and Technology Innovation Program of Hunan Province under Grant No. 2024RC1050.

\bibliography{references}

\end{document}